\documentclass[12pt]{article}

\usepackage{amsmath}
\usepackage{amssymb}
\usepackage{amsfonts}
\usepackage{latexsym}
\usepackage{color}

\catcode `\@=11 \@addtoreset{equation}{section}

\catcode `\@=12

\newcommand{\be}{\begin{equation}}
\newcommand{\en}{\end{equation}}
\newcommand{\bea}{\begin{eqnarray}}
\newcommand{\ena}{\end{eqnarray}}
\newcommand{\beano}{\begin{eqnarray*}}
\newcommand{\enano}{\end{eqnarray*}}
\newcommand{\bee}{\begin{enumerate}}
\newcommand{\ene}{\end{enumerate}}

\newcommand{\mc}{\mathcal}

\newcommand{\D}{{\mc D}}

\newcommand{\Sc}{{\cal S}}
\newcommand{\E}{{\cal E}}
\newcommand{\F}{{\cal F}}
\newcommand{\G}{{\cal G}}

\newcommand{\Lc}{{\cal L}}

\newcommand{\1}{1 \!\! 1}

\newcommand{\Hil}{\mc H}

\newtheorem{thm}{Theorem}

\newtheorem{prop}[thm]{Proposition}
\newtheorem{defn}[thm]{Definition}

\catcode `\@=11 \@addtoreset{equation}{section}
\catcode `\@=12

\begin{document}

\thispagestyle{empty}

\vspace*{2cm}

\begin{center}
{\Large \bf $\D$ pseudo-bosons in quantum models}   \vspace{2cm}\\

{\large F. Bagarello}\\
  Dipartimento di Energia, Ingegneria dell'Informazione e Modelli Matematici,\\
Facolt\`a di Ingegneria, Universit\`a di Palermo,\\ I-90128  Palermo, Italy\\
e-mail: fabio.bagarello@unipa.it\\
Tel: +390912389722; Fax: +39091427258\\
home page: www.unipa.it/fabio.bagarello

\vspace{.4cm}

 {\large M. Lattuca}\\
  Dipartimento di Energia, Ingegneria dell'Informazione e Modelli Matematici,\\
Facolt\`a di Ingegneria, Universit\`a di Palermo,\\ I-90128  Palermo, Italy\\
e-mail: marghylt@libero.it

\end{center}

\vspace*{2cm}

\begin{abstract}
\noindent We show how some recent models of PT-quantum mechanics perfectly fit into the settings of $\D$ pseudo-bosons, as introduced by one of
us. Among the others, we also consider a model of non-commutative quantum mechanics, and we show that this model too can be described in terms
of $\D$ pseudo-bosons.

\end{abstract}

\vspace{2cm}


\vfill


\newpage

\section{Introduction}

In a series of papers, \cite{bagpb1}-\cite{bagrev}, one of us (FB) considered two operators $a$ and $b$, with $b\neq a^\dagger$, acting on a
Hilbert space $\Hil$, and satisfying the  commutation rule $[a,b]=\1$. Under suitable assumptions, a nice functional structure has been
deduced, and some connections with physics, and in particular with quasi-hermitian (or PT) quantum mechanics\footnote{or variations on the same
scheme.} and with the technique of intertwining operators, have been established. The particle-like excitations associated to this structure
have been called {\em pseudo-bosons} (PB). The assumptions used in that construction have been checked for a series of (quantum mechanical)
models.

More recently, \cite{bagnewpb}, FB has introduced a slightly different version os PB, the so-called $\D$-PB, for which all those mathematical
{\em dangerous} aspects, related to the fact that the operators involved are usually unbounded, can be discussed in a more appropriate
settings.

This paper continues a series of other papers, \cite{bagpb4} and \cite{bagnewpb} among the others, whose aim is to show that pseudo-bosons or
$\D$-PB are indeed rather frequent in the literature on PT-quantum mechanics, and may work very well as an unifying framework, at least for
those hamiltonians whose eigenvalues are linear in the quantum numbers. In particular, the models considered here were first introduced and
analyzed, under a similar point of view, in \cite{ben1,miao1,miao2}. The original interest in these (and similar) models arose mainly because
of the possibility of having explicit hamiltonians, manifestly non-selfadjoint, which possess only real eigenvalues. In this perspective, and
also in view of recent studies on gain-loss systems, \cite{circu}, hamiltonians of this kind have attracted a big interest in the physicists
community, both from a theoretical and from an experimental point of view. Therefore, a deeper understanding of these system is surely
important for further developments of these aspects of quantum mechanics. In particular, in this article we show that models, which were
originally introduced in connection with PT-quantum mechanics, could be quite naturally analyzed in terms of $\D$-PB, making explicit  the
reason why the eigenvalues of their hamiltonians are indeed real numbers.

This article is organized as follows: in the next section we review our definition of $\D$-PB, that is of those PB which are, somehow,
associated to a certain subspace $\D$, dense in the Hilbert space $\Hil$ on which our operators $a$ and $b$ act. For much more details we refer
to \cite{bagnewpb}. Sections III, IV and V contains our examples, while our conclusions are given in Section VI.

\section{$\D$ pseudo-bosons}

Let $\Hil$ be a given Hilbert space with scalar product $\left<.,.\right>$ and related norm $\|.\|$. Let further $a$ and $b$ be two operators
on $\Hil$, with domains $D(a)$ and $D(b)$ respectively, $a^\dagger$ and $b^\dagger$ their adjoint, and let $\D$ be a dense subspace of $\Hil$
such that $a^\sharp\D\subseteq\D$ and $b^\sharp\D\subseteq\D$, where $x^\sharp$ is $x$ or $x^\dagger$. Incidentally,it may be worth noticing
that we are not requiring here that $\D$ coincides with, e.g. $D(a)$ or $D(b)$. Nevertheless, for obvious reasons, $\D\subseteq D(a^\sharp)$
and $\D\subseteq D(b^\sharp)$.

\begin{defn}\label{def21}
The operators $(a,b)$ are $\D$-pseudo bosonic ($\D$-pb) if, for all $f\in\D$, we have
\be
a\,b\,f-b\,a\,f=f.
\label{31}\en
\end{defn}
 Sometimes, to simplify the notation, instead of (\ref{31}) we will simply write $[a,b]=\1$, having in mind that both sides of this equation
have to act on $f\in\D$.

\vspace{2mm}

Our first working assumptions are the following:

\vspace{2mm}

{\bf Assumption $\D$-pb 1.--}  there exists a non-zero $\varphi_{ 0}\in\D$ such that $a\varphi_{ 0}=0$.

\vspace{1mm}

{\bf Assumption $\D$-pb 2.--}  there exists a non-zero $\Psi_{ 0}\in\D$ such that $b^\dagger\Psi_{ 0}=0$.

\vspace{2mm}

Then, if $(a,b)$ satisfy Definition \ref{def21}, it is obvious that $\varphi_0\in D^\infty(b):=\cap_{k\geq0}D(b^k)$ and that $\Psi_0\in
D^\infty(a^\dagger):=\cap_{k\geq0}D((a^\dagger)^k)$, so that the vectors \be \varphi_n:=\frac{1}{\sqrt{n!}}\,b^n\varphi_0,\qquad
\Psi_n:=\frac{1}{\sqrt{n!}}\,{a^\dagger}^n\Psi_0, \label{32}\en $n\geq0$, can be defined and they all belong to $\D$. We introduce, as in
\cite{bagnewpb}, $\F_\Psi=\{\Psi_{ n}, \,n\geq0\}$ and $\F_\varphi=\{\varphi_{ n}, \,n\geq0\}$. Once again, since $\D$ is stable under the
action of $a^\sharp$ and $b^\sharp$, we deduce that each $\varphi_n$ and each $\Psi_n$ belongs to the domains of $a^\sharp$, $b^\sharp$ and
$N^\sharp$, where $N=ba$.

It is now simple to deduce the following lowering and raising relations:
\be
\left\{
    \begin{array}{ll}
b\,\varphi_n=\sqrt{n+1}\varphi_{n+1}, \qquad\qquad\quad\,\, n\geq 0,\\
a\,\varphi_0=0,\quad a\varphi_n=\sqrt{n}\,\varphi_{n-1}, \qquad\,\, n\geq 1,\\
a^\dagger\Psi_n=\sqrt{n+1}\Psi_{n+1}, \qquad\qquad\quad\, n\geq 0,\\
b^\dagger\Psi_0=0,\quad b^\dagger\Psi_n=\sqrt{n}\,\Psi_{n-1}, \qquad n\geq 1,\\
       \end{array}
        \right.
\label{33}\en as well as the following eigenvalue equations: $N\varphi_n=n\varphi_n$ and  $N^\dagger\Psi_n=n\Psi_n$, $n\geq0$. As a consequence
of these  equations,  choosing the normalization of $\varphi_0$ and $\Psi_0$ in such a way $\left<\varphi_0,\Psi_0\right>=1$, we deduce that
\be \left<\varphi_n,\Psi_m\right>=\delta_{n,m}, \label{34}\en
 for all $n, m\geq0$. The third assumption is the following:

\vspace{2mm}

{\bf Assumption $\D$-pb 3.--}  $\F_\varphi$ is a basis for $\Hil$.

\vspace{1mm}

This assumption introduces, apparently, an asymmetry  between $\F_\varphi$ and $\F_\Psi$, since this last is not required to be a basis as
well. However, under the above assumptions, we can check that $\F_\varphi$ is a basis for $\Hil$ if and only if $\F_\Psi$ is also a basis for
$\Hil$, \cite{bagnewpb}. Moreover, if $\F_\varphi$ and $\F_\Psi$ are Riesz basis for $\Hil$, we  call our $\D$-PB {\em regular}, as we have
done in our previous papers.

\vspace{2mm}

{\bf Remarks:--} (1) As it is widely discussed in, e.g., \cite{bagpb1}, $\F_\varphi$ and $\F_\Psi$ are Riesz bases if and only if the so-called
metric operator, which could be formally written as $S_\varphi=\sum|\varphi_n\left>\right<\varphi_n|$, is bounded with bounded inverse. Since
$S_\varphi$ is also positive, this would allow us to define a different but equivalent scalar product in $\Hil$, with respect to which $N$
becomes self-adjoint. When $S_\varphi$ is not bounded, i.e. when $\F_\varphi$ and $\F_\Psi$ are not Riesz bases, this possibility is forbidden.

(2) It might be worth noticing that requiring that $\F_\varphi$ to be a basis is much more, for non o.n. sets, than requiring $\F_\varphi$ to
be complete. Counterexamples can be found in \cite{bagnewpb,heil}.

\vspace{2mm}

A weaker version of Assumption $\D$-pb 3 was also introduced in \cite{bagnewpb}:

\vspace{2mm}

{\bf Assumption $\D$-pbw 3.--}  $\F_\varphi$ and $\F_\Psi$ are $\G$-quasi bases for $\Hil$.

\vspace{2mm}

This means that a dense subspace $\G\subset\Hil$ exists such that $\varphi_n, \Psi_n\in\G$ and
$$
\left<f,g\right>=\sum_{n\geq0}\left<f,\eta_n\right>\left<\Phi_n,g\right>=\sum_{n\geq0}\left<f,\Phi_n\right>\left<\eta_n,g\right>,
$$
for all $f, g\in\G$. Then we have a {\em weak} resolution of the identity.

\subsection{$\D$-conjugate operators}

In this section we slightly refine the structure.

We start considering a self-adjoint, invertible, operator $\Theta$, which leaves,  together with $\Theta^{-1}$, $\D$ invariant:
$\Theta\D\subseteq\D$, $\Theta^{-1}\D\subseteq\D$. Then, if Assumptions $\D$-pb 1, 2 and 3 hold, we can introduce the following definition:

\begin{defn}\label{def41}
We will say that $(a,b^\dagger)$ are $\Theta-$conjugate if $af=\Theta^{-1}b^\dagger\,\Theta\,f$, for all $f\in\D$.
\end{defn}
Briefly, we will often write $a=\Theta^{-1}b^\dagger\,\Theta$. In \cite{bagnewpb} it is shown, for instance, that $(a,b^\dagger)$ are
$\Theta-$conjugate if and only if $(b,a^\dagger)$ are $\Theta-$conjugate. It is also shown that we can always assume that
$\left<\varphi_0,\Theta\varphi_0\right>=1$, at least if $\varphi_0\notin \ker(\Theta)$, and that the operators $(a,b^\dagger)$ are
$\Theta-$conjugate if and only if $\Psi_n=\Theta\varphi_n$, for all $n\geq0$. This result is particularly interesting, since gives necessary
and sufficient conditions for $\F_\varphi$ and $\F_\Psi$ to be related by a certain operator, which plays a crucial role in all our framework.

When $(a,b^\dagger)$ are $\Theta-$conjugate then (i) $\left<f,\Theta f\right>>0$ for all non zero $f\in D(\Theta)$ and (ii) $
Nf=\Theta^{-1}N^\dagger\Theta f$ for all $f\in \D$, so that $N$ is a  strongly crypto-hermitian operator

We end this introductive section by stating the following  result, again contained in \cite{bagnewpb}:

let $\E=\{e_n\in\Hil, n\geq0\}$ be an o.n. basis of $\Hil$ and let us consider a self-adjoint, invertible operator $T$, such that $e_n\in
D(T)\cap D(T^{-1})$ for all $n$.  Then the vectors  $c_n=Te_n$ and $d_n=T^{-1}e_n$, $n\geq0$, are well defined in $\Hil$. We call
$\F_c=\{c_n,\,n\geq0\}$ and $\F_d=\{d_n,\,n\geq0\}$.

\begin{prop}\label{propa1} Under the above assumptions:
 (i) the sets $\F_c$ and $\F_d$ are biorthogonal; (ii) if $f\in D(T)$ is orthogonal to all the $c_n$,
 then $f=0$; (iii) if $f\in D(T^{-1})$ is orthogonal to all the $d_n$, then $f=0$; (iv)  $\F_c$  and $\F_d$ are $D(T)\cap D(T^{-1})$-quasi bases.
  \end{prop}

The outcome of this proposition is that we don't really need $\F_c$ and $\F_d$ to be  Riesz bases in order to get some sort of resolution of the identity. This is possible also if $T$ or $T^{-1}$, or both, are unbounded, at least when Proposition
\ref{propa1} applies. Of course, when both $T$ and $T^{-1}$ are bounded, then $\F_c$ and $\F_d$ are Riesz bases.

\section{Example one}

The first example we want to consider here was originally introduced in \cite{ben1} and then considered further in \cite{miao1}. The starting
point is the following, manifestly non self-adjoint, hamiltonian: \be H=(p_1^2+x_1^2)+(p_2^2+x_2^2+2ix_2)+2\epsilon x_1x_2,\label{41}\en where
$\epsilon$ is a real constant, with $\epsilon\in]-1,1[$. Here the following commutation rules are assumed: $[x_j,p_k]=i\delta_{j,k}\1$, $\1$
being the identity operator on $\Lc^2({\Bbb R}^2)$. All the other commutators are zero.

Repeating the same steps as in \cite{miao1}, we can perform some changes of variables which allow us to write the hamiltonian in a different,
and more convenient, form:
\begin{enumerate}

\item first of all we introduce the {\em capital} operators $P_j$, $X_j$, $j=1,2$, via
$$
P_1:=\frac{1}{2a}(p_1+\xi p_2),\quad P_2:=\frac{1}{2b}(p_1-\xi p_2),\quad X_1:=a(x_1+\xi x_2),\quad X_2:=b(x_1-\xi x_2),
$$
where $\xi$ can be $\pm 1$, while $a$ and $b$ are real, non zero, arbitrary constants. These operators satisfy the same canonical
commutation rules as the original ones:  $[X_j,P_k]=i\delta_{j,k}\1$.
\item Secondly, we introduce the operators
$$
\Pi_1=P_1,\quad \Pi_2=P_2,\quad q_1=X_1+i\frac{a\xi}{1+\epsilon\, \xi}, \quad q_2=X_2-i\frac{b\xi}{1-\epsilon\, \xi}.
$$
The first clear fact is that $\Pi_j^\dagger=\Pi_j$, while $q_j^\dagger\neq q_j$, $j=1,2$. However, the commutation rules are preserved: $[q_j,\Pi_k]=i\delta_{j,k}\1$.
\item The third step consists in introducing the following operators:
\be
a_1=\frac{a}{\sqrt[4]{1+\epsilon\, \xi}}\left(i\Pi_1+\frac{\sqrt{1+\epsilon\, \xi}}{2a^2}\,q_1\right),\quad a_2=\frac{a}{\sqrt[4]{1-\epsilon\, \xi}}\left(i\Pi_2+\frac{\sqrt{1-\epsilon\, \xi}}{2b^2}\,q_2\right),
\label{42}\en
and
\be
b_1=\frac{a}{\sqrt[4]{1+\epsilon\, \xi}}\left(-i\Pi_1+\frac{\sqrt{1+\epsilon\, \xi}}{2a^2}\,q_1\right),\quad b_2=\frac{a}{\sqrt[4]{1-\epsilon\, \xi}}\left(-i\Pi_2+\frac{\sqrt{1-\epsilon\, \xi}}{2b^2}\,q_2\right).
\label{43}\en
It may be worth remarking that $b_j\neq a_j^\dagger$, the reason being that $q_j$ are not self-adjoint.
These operators satisfy the pseudo-bosonic commutation rules
\be
[a_j,b_k]=\delta_{j,k}\1,
\label{44}\en
the other commutators being zero.

\end{enumerate}

Going back to $H$, and introducing the operators $N_j:=b_ja_j$, we can write \be H=H_1+H_2+\frac{1}{1-\epsilon^2}\,\1,\qquad
H_1=\sqrt{1+\epsilon\, \xi}(2N_1+\1), \quad H_2=\sqrt{1-\epsilon\, \xi}(2N_2+\1). \label{45}\en These results are essentially already contained
in \cite{miao1}, even if not exactly in this form. Our next step consists in checking if the two-dimensional version of the general framework
described in Section II applies to the present model. In other words, we want to check if Assumptions $\D$-pb 1, $\D$-pb 2 and $\D$-pb 3 hold
true or not in $\Hil=\Lc^2({\Bbb R}^2)$.

For that, the first thing to do is to rewrite the operators $a_j$ and $b_j$ in terms of the original $x_j$ and $p_j$, used in (\ref{41}): $$
\left\{
    \begin{array}{ll}
a_1=\frac{1}{2\sqrt[4]{1+\epsilon\, \xi}}\left((ip_1+\sqrt{1+\epsilon\, \xi}\,x_1)+\xi(ip_2+\sqrt{1+\epsilon\, \xi}\,x_2)+
i\,\frac{\xi}{\sqrt{1+\epsilon\, \xi}}\right),\\
a_2=\frac{1}{2\sqrt[4]{1-\epsilon\, \xi}}\left((ip_1+\sqrt{1-\epsilon\, \xi}\,x_1)-\xi(ip_2+\sqrt{1-\epsilon\, \xi}\,x_2)-
i\,\frac{\xi}{\sqrt{1-\epsilon\, \xi}}\right),\\
b_1=\frac{1}{2\sqrt[4]{1+\epsilon\, \xi}}\left((-ip_1+\sqrt{1+\epsilon\, \xi}\,x_1)+\xi(-ip_2+\sqrt{1+\epsilon\, \xi}\,x_2)+
i\,\frac{\xi}{\sqrt{1+\epsilon\, \xi}}\right),\\
b_2=\frac{1}{2\sqrt[4]{1-\epsilon\, \xi}}\left((-ip_1+\sqrt{1-\epsilon\, \xi}\,x_1)-\xi(-ip_2+\sqrt{1-\epsilon\, \xi}\,x_2)-
i\,\frac{\xi}{\sqrt{1-\epsilon\, \xi}}\right).
       \end{array}
        \right.
$$
We now have to find a dense subspace $\D$ of $\Lc^2({\Bbb R}^2)$ which is stable under the action of $a_j$, $b_j$ and their adjoints. Moreover
 $\D$ must also contains the two vacua of $a_j$ and $b_j^\dagger$, if they exist. Hence, from a practical point of view, it is convenient to look
 first for a solution of the equations $a_1\varphi_{0,0}(x_1,x_2)=a_2\varphi_{0,0}(x_1,x_2)=0$ and
 $b_1^\dagger\Psi_{0,0}(x_1,x_2)=b_2^\dagger\Psi_{0,0}(x_1,x_2)=0$. Using $p_j=-i\frac{\partial}{\partial x_j}$, these are simple two-dimensional
  differential equations which can be easily solved, and the results are
\be
\left\{
    \begin{array}{ll}
\varphi_{0,0}(x_1,x_2)=N\exp\left\{-\frac{1}{2}\alpha_+(x_1^2+x_2^2)-k_-x_1-k_+x_2-\xi\alpha_-x_1x_2\right\},\\
\Psi_{0,0}(x_1,x_2)=N'\exp\left\{-\frac{1}{2}\alpha_+(x_1^2+x_2^2)+k_-x_1+k_+x_2-\xi\alpha_-x_1x_2\right\},
\end{array}
        \right.
        \label{46}\en
where we have introduced the following constants:
$$
\alpha_{\pm}=\frac{1}{2}\left(\sqrt{1+\epsilon\, \xi}\pm \sqrt{1-\epsilon\, \xi}\right),\quad
k_-=\frac{-i\xi\alpha_-}{\sqrt{1-\epsilon^2}},\quad k_+=\frac{i\alpha_+}{\sqrt{1-\epsilon^2}}.
$$
$N$ and $N'$ in (\ref{46}) are normalization constants, fixed by the requirement that $\left<\varphi_{0,0},\Psi_{0,0}\right>=1$. This is possible,
since we can easily check that $\varphi_{0,0}(x_1,x_2), \Psi_{0,0}(x_1,x_2)\in \Lc^2({\Bbb R}^2)$. As a matter of fact, there is more than this:
both $\varphi_{0,0}(x_1,x_2)$ and $\Psi_{0,0}(x_1,x_2)$ belong to $\Sc({\Bbb R}^2)$, the set of those $C^\infty$ functions which decrease to zero,
together
with their derivatives, faster than any inverse power of $x_1$ and $x_2$. Since $\Sc({\Bbb R}^2)$ is dense in $\Lc^2({\Bbb R}^2)$, it is natural to
identify $\D$ with $\Sc({\Bbb R}^2)$. This is a good choice. In fact, other than having $\varphi_{0,0}(x_1,x_2),\,\Psi_{0,0}(x_1,x_2)\in\D$,
  $\D$ is also stable under the action of $a_j$, $b_j$ and of their adjoints.

At this point we can construct the new functions
$\varphi_{n_1,n_2}(x_1,x_2)=\frac{1}{\sqrt{n_1!\,n_2!}}b_1^{n_1}b_2^{n_2}\varphi_{0,0}(x_1,x_2)$ and
$\Psi_{n_1,n_2}(x_1,x_2)=\frac{1}{\sqrt{n_1!\,n_2!}}{a_1^\dagger}^{n_1}{a_2^\dagger}^{n_2}\Psi_{0,0}(x_1,x_2)$, and the related sets
$\F_\varphi=\{\varphi_{n_1,n_2}(x_1,x_2), n_j\geq0\}$, $\F_\Psi=\{\Psi_{n_1,n_2}(x_1,x_2), n_j\geq0\}$. It is clear that both
$\varphi_{n_1,n_2}(x_1,x_2)$ and $\Psi_{n_1,n_2}(x_1,x_2)$ differ from $\varphi_{0,0}(x_1,x_2)$ and $\Psi_{0,0}(x_1,x_2)$  for some polynomial
in $x_1$ and $x_2$. Hence they are still functions in $\Sc({\Bbb R}^2)$, as expected.

\vspace{2mm} The final effort consists now in proving that $\F_\varphi$ and $\F_\Psi$ are bases for $\Hil$. This is not evident, in principle.
What is much easier to check is that these sets are both complete in $\Hil$, but we know that completeness of a certain set does not imply that
that set is a basis. Following \cite{ben1} we define an unbounded, self-adjoint and invertible operator
$T=e^{\frac{1}{1-\epsilon^2}(p_2-\epsilon p_1)}$. Then, simple computations show that \be T\,H\,T^{-1}=(p_1^2+x_1^2)+(p_2^2+x_2^2)+2\epsilon
x_1x_2+\frac{1}{1-\epsilon^2}=:h.\label{47}\en It is clear that, contrarily to $H$, $h=h^\dagger$. For $h$ we can repeat essentially the same
procedure as before. In particular, we can again introduce the capital operators $P_j$, $X_j$ as before, and the operators $$
A_1=\frac{a}{\sqrt[4]{1+\epsilon\, \xi}}\left(i\Pi_1+\frac{\sqrt{1+\epsilon\, \xi}}{2a^2}\,X_1\right),\quad A_2=\frac{b}{\sqrt[4]{1-\epsilon\,
\xi}}\left(i\Pi_2+\frac{\sqrt{1-\epsilon\, \xi}}{2b^2}\,X_2\right),$$ and the adjoints $A_j^\dagger$. These are true bosonic operators:
$[A_j,A_k^\dagger]=\delta_{j,k}\1$, in terms of which $h=h_1+h_2+\frac{1}{1-\epsilon^2}\,\1$, with $h_1=\sqrt{1+\epsilon\, \xi}(2\hat N_1+\1)$
and $h_2=\sqrt{1-\epsilon\, \xi}(2\hat N_2+\1)$, where $\hat N_j:=A_j^\dagger A_j$ is a bosonic number operator.

Now, if $\Phi_{0,0}$ is the vacuum of $A_j$, $A_1\Phi_{0,0}=A_2\Phi_{0,0}=0$, we can construct, {\em more solito}, the set
$\F_\Phi:=\{\Phi_{n_1,n_2}, \,n_j\geq0\}$, where $\Phi_{n_1,n_2}=\frac{1}{\sqrt{n_1!\,n_2!}}{A_1^\dagger}^{n_1}{A_2^\dagger}^{n_2}\Phi_{0,0}$.
$\F_\Phi$ is an o.n. basis for $\Hil$, and the $\Phi_{n_1,n_2}(x_1,x_2)$ can be factorized as follows:
$$
\Phi_{n_1,n_2}(x_1,x_2)=\Phi_{n_1}(x_1)\Phi_{n_2}(x_2),
$$
where $\Phi_n(x)$ are the usual eigenstates of a one-dimensional harmonic oscillator: $\Phi_n(x)=\frac{1}{\sqrt{2^n\pi
n!}}H_n(x)e^{-\frac{1}{2} x^2}$. Each $\Phi_{n_1,n_2}(x_1,x_2)$ belongs to $\D$. Moreover, $\Phi_{n_1,n_2}\in D(T)\cap D(T^{-1})$. In
particular we can check that $T\Phi_{n_1,n_2}(x_1,x_2)=\Phi_{n_1,n_2}(x_1+\delta_1,x_2+\delta_2)$, where
$\delta_1=\frac{i}{1-\epsilon^2}(a\epsilon-\xi)$ and $\delta_2=\frac{i}{1-\epsilon^2}(b\epsilon+\xi)$. Needless to say,
$T^{-1}\Phi_{n_1,n_2}(x_1,x_2)=\Phi_{n_1,n_2}(x_1-\delta_1,x_2-\delta_2)$.

It is now possible to check that, for all $n_1$ and $n_2$, $T\Phi_{n_1,n_2}(x_1,x_2)=\Psi_{n_1,n_2}(x_1,x_2)$ and
$T^{-1}\Phi_{n_1,n_2}(x_1,x_2)=\varphi_{n_1,n_2}(x_1,x_2)$. For that it is convenient to recall that the following equations must all be
satisfied: $h\Phi_{n_1,n_2}=E_{n_1,n_2}\Phi_{n_1,n_2}$, $H\varphi_{n_1,n_2}=E_{n_1,n_2}\varphi_{n_1,n_2}$, $THT^{-1}=h$, as well as
$E_{n_1,n_2}=\sqrt{1+\epsilon\, \xi}(2n_1+1)+\sqrt{1-\epsilon\, \xi}(2n_2+1)+\frac{1}{1-\epsilon^2}$. If $\epsilon\neq0$, each $E_{n_1,n_2}$ is
not degenerate. It is convenient here to work in this assumption, even because, if $\epsilon=0$, the original hamiltonian $H$ simplifies a lot
and becomes less interesting for us. Since  $\Phi_{n_1,n_2}\in D(T)$, equation $h\Phi_{n_1,n_2}=E_{n_1,n_2}\Phi_{n_1,n_2}$ can be rewritten as
follows: $H(T^{-1}\Phi_{n_1,n_2})=E_{n_1,n_2}(T^{-1}\Phi_{n_1,n_2})$. Therefore $T^{-1}\Phi_{n_1,n_2}$ must be proportional to
$\varphi_{n_1,n_2}$. For similar reasons, we can check that $T\Phi_{n_1,n_2}$ must be proportional to $\Psi_{n_1,n_2}$, since $H^\dagger
\Psi_{n_1,n_2}=E_{n_1,n_2}\Psi_{n_1,n_2}$. These proportionality constants can be taken all equal to one. We are in the conditions of
Proposition \ref{propa1}; therefore $\F_\varphi$ and $\F_\Psi$ are both $D(T)\cap D(T^{-1})$-quasi bases for $\Hil$.  This means that
Assumption $\D$-pbw 3 is also satisfied.

Let us now take $\Theta:=T^2$. It is clear that $\Theta^{-1}$ exists and that, together with $\Theta$, leaves $\D$ invariant. Moreover
$\Psi_{n_1,n_2}=\Theta\varphi_{n_1,n_2}$ so that, as discussed in Section II, $(a_j,b_j^\dagger)$ turn out to be $\Theta$-conjugate. The
intertwining relation  $N_j f=\Theta^{-1}N_j^\dagger\Theta f$, $f\in\D$, holds true. Formally, we can write $\Theta=\sum_{\bf
k}|\Psi_{k_1,k_2}\left>\right<\Psi_{k_1,k_2}|$ and $\Theta^{-1}=\sum_{\bf k}|\varphi_{k_1,k_2}\left>\right<\varphi_{k_1,k_2}|$. It is clear
that these series cannot be uniformly convergent, since both $\Theta$ and $\Theta^{-1}$ are unbounded.

\section{Example two}

In this section we consider a different quantum mechanical model, originally introduced, in our knowledge, in \cite{miao2}. The starting point
is the following manifestly non hermitian hamiltonian, \be
H=\frac{1}{2}(p_1^2+x_1^2)+\frac{1}{2}(p_2^2+x_2^2)+i\left[A(x_1+x_2)+B(p_1+p_2)\right], \label{51}\en where $A$ and $B$ are real constants,
while $x_j$ and $p_j$ are the  self-adjoint position and momentum operators, satisfying $[x_j,p_k]=i\delta_{j,k}\1$.

As in the previous example, we can introduce new variables to write $H$ is a different, more convenient, form. For that we first put
$$
P_1=p_1+iB,\quad P_2=p_2+iB,\quad X_1=x_1+iA,\quad X_2=x_2+iA,
$$
and then \be a_j=\frac{1}{\sqrt{2}}(X_j+iP_j),\qquad  b_j=\frac{1}{\sqrt{2}}(X_j-iP_j),\label{52}\en $j=1,2$. It is easy to check that
$[X_j,P_k]=i\delta_{j,k}\1$, $[a_j,b_k]=\delta_{j,k}\1$, and that, since $X_j^\dagger\neq X_j$ and $P_j^\dagger\neq P_j$, $b_j\neq
a_j^\dagger$. Introducing further $N_j=b_ja_j$ we can rewrite $H$ as follows: $H=N_1+N_2+(A^2+B^2+1)\1$.

The eigenstates of $H$ and $H^\dagger$ can now be easily constructed if assumptions $\D$-pb 1 and $\D$-pb 2 are satisfied. If assumption
$\D$-pb 3 is also satisfied, then the sets of their eigenstates are biorthogonal bases for $\Hil=\Lc^2({\Bbb R}^2)$.

To check that all these steps can be carried out, we proceed as before, writing first $a_j$ and $b_j$ in terms of the original operators $x_j$
and $p_j$. In this case the procedure is quite easy, and we find
$$
a_j=\frac{1}{\sqrt{2}}(x_j+ip_j+C),\qquad b_j=\frac{1}{\sqrt{2}}(x_j-ip_j+D),
$$
$j=1,2$, where $C=iA-B$ and $D=iA+B$. The two vacua of $a_j$ and $b_j^\dagger$ are respectively
$$
\varphi_{0,0}(x_1,x_2)=Ne^{-\frac{1}{2}(x_1^2+x_2^2)-C(x_1+x_2)}, \quad \Psi_{0,0}(x_1,x_2)=N'e^{-\frac{1}{2}(x_1^2+x_2^2)-\overline{D} (x_1+x_2)},
$$
where $N$ and $N'$ are normalization constant chosen in such a way $\left<\varphi_{0,0},\Psi_{0,0}\right>=1$. Also for this example we observe
that both $\varphi_{0,0}(x_1,x_2)$ and $\Psi_{0,0}(x_1,x_2)$ belong to $\Sc({\Bbb R}^2)$, which we take as the space $\D$ for our PB. Due to
the particularly easy expressions for, say, $b_j$ and $\varphi_{0,0}(x_1,x_2)$, it is easy to see that $\varphi_{n_1,n_2}(x_1,x_2)$ can be
factorized. In fact we have
$$
\varphi_{n_1,n_2}(x_1,x_2)=\frac{N}{\sqrt{n_1!n_2!2^{n_1+n_2}}}\,\left[\left(x_1-\frac{\partial}{\partial x_1}+D\right)^{n_1}
e^{-\frac{1}{2}x_1^2-Cx_1}\right]\left[\left(x_2-\frac{\partial}{\partial x_2}+D\right)^{n_2}
e^{-\frac{1}{2}x_2^2-Cx_2}\right],
$$
while $\Psi_{n_1,n_2}(x_1,x_2)$ can be deduced from $\varphi_{n_1,n_2}(x_1,x_2)$ simply replacing $C$ with $\overline{D}$ and viceversa
everywhere. Incidentally we observe that, as expected, $\varphi_{n_1,n_2}(x_1,x_2)$ and $\Psi_{n_1,n_2}(x_1,x_2)$ are all in $\Sc({\Bbb R}^2)$.

The hard part of the job is now the proof that both $\F_\varphi=\{\varphi_{n_1,n_2}(x_1,x_2), n_j\geq0\}$, $\F_\Psi=\{\Psi_{n_1,n_2}(x_1,x_2),
n_j\geq0\}$ are bases for $\Hil$. Again, we will show that Proposition \ref{propa1} is useful to this task. In fact, let us introduce the
following unbounded, self-adjoint, invertible operator $T$:
$$
T=e^{-A(p_1+p_2)+B(x_1+x_2)}.
$$
It is possible to see that $H=T\tilde hT^{-1}$, where $\tilde h=\frac{1}{2}(p_1^2+x_1^2)+\frac{1}{2}(p_2^2+x_2^2)+(A^2+B^2)\1$. Therefore, if
we introduce the standard bosonic operators $c_j=\frac{1}{\sqrt{2}}(x_j+ip_j)$, together with their adjoints, we see that $\tilde h=c_1^\dagger
c_1+c_2^\dagger c_2+(A^2+B^2+1)\1$. The eigenvalues of $\tilde h$ are $E_{n_1,n_2}=n_1+n_2+A^2+B^2+1$, and the related eigenvectors are
constructed as usual, for a two-dimensional harmonic oscillator: given $\Phi_{0,0}(x_1,x_2)\in \Hil$ such that $c_j\Phi_{0,0}=0$, $j=1,2$, the
set of eigenstates of $\tilde h$ are obtained using the raising operators:
$\Phi_{n_1,n_2}:=\frac{1}{\sqrt{n_1!n_2!}}(c_1^\dagger)^{n_1}(c_2^\dagger)^{n_2}\Phi_{0,0}$, $n_j\geq0$. The set $\F_\Phi=\{\Phi_{n_1,n_2},\,
n_j\geq0\}$ is an o.n. basis for $\Hil$, and it is a simple exercise to check that, not only  $\Phi_{n_1,n_2}\in D(T)\cap D(T^{-1})$, but also
that $\varphi_{n_1,n_2}=T\Phi_{n_1,n_2}$ and $\Psi_{n_1,n_2}=T^{-1}\Phi_{n_1,n_2}$. We are in the conditions of Proposition \ref{propa1}, so
that our claim follows: $\F_\varphi$ and $\F_\Psi$ are $D(T)\cap D(T^{-1})$-quasi bases for $\Hil$.

The operator $\Theta$ is now $\Theta=T^{-2}=e^{2A(p_1+p_2)-2B(x_1+x_2)}$, which maps $\D$ into itself. The same final considerations as in
Example one can be repeated also here.

\section{Example three}

The third example we want to consider here is a noncommutative version of the previous one, in which the hamiltonian looks formally as that in
(\ref{51}): \be \hat H=\frac{1}{2}(\hat p_1^2+\hat x_1^2)+\frac{1}{2}(\hat p_2^2+\hat x_2^2)+i\left[A(\hat x_1+\hat x_2)+B(\hat p_1+\hat
p_2)\right], \label{61}\en where again $A$ and $B$ are real constants. The difference is that the self-adjoint operators $\hat x_j$ and $\hat
p_k$ are now assumed to satisfy the following commutation rules: \be [\hat x_j,\hat p_k]=i\delta_{j,k}\1,\quad [\hat x_j,\hat
x_k]=i\theta\epsilon_{j,k}\1,\quad [\hat p_j,\hat p_k]=i\tilde\theta\epsilon_{j,k}\1.\label{62}\en Here $\theta$ and $\tilde\theta$ are two
small parameters, which measure the noncommutativity of the system, and we have $\epsilon_{j,j}=0$, $\epsilon_{1,2}=-\epsilon_{2,1}=1$.

Following \cite{miao2}, we will set up a perturbative approach for (the first part of) this example. In particular, in what follows we will
only keep the terms which are, at most, linear in $\theta$ and $\tilde\theta$, neglecting all the quadratic, cubic, ...terms. It may be
interesting to notice also that, in some papers on noncommutative quantum mechanics, see \cite{abg2} and references therein, $\tilde\theta$ is
taken to be zero and the noncommutative aspects are contained only in the position, and not in the momentum, operators.

With this in mind, if we introduce two pairs of canonically conjugate operators, $(x_j,p_j)$, $j=1,2$\footnote{i.e.
$[x_j,p_k]=i\delta_{j,k}\1$, $[x_j,x_k]=[p_j,p_k]=0$.}, we can recover (\ref{62}) if we assume that \be \hat x_1=x_1-\frac{1}{2}\theta
p_2,\quad\hat x_2=x_2+\frac{1}{2}\theta p_1,\quad \hat p_1=p_1+\frac{1}{2}\tilde\theta x_2,\quad \hat p_2=p_2-\frac{1}{2}\tilde\theta x_1.
\label{63}\en Then $\hat H$ can be rewritten, up to corrections quadratic in $\theta$ and $\tilde\theta$, as
$$
\hat H=\frac{1}{2}(p_1^2+ x_1^2)+\frac{1}{2}( p_2^2+ x_2^2)+i\left[A( x_1+ x_2)+B( p_1+p_2)\right]+\frac{1}{2}(\theta+\tilde\theta)(p_1x_2-p_2x_1)+
$$
\be\label{64} +i\left[\frac{A\theta}{2}(p_1-p_2)-\frac{B\tilde\theta}{2}(x_1-x_2)\right] \en Defining now new, non self-adjoint, operators
$P_j=p_j+iB_j$, $X_j=x_j+iA_j$, $j=1,2$, we observe that $[X_j,P_k]=i\delta_{j,k}\1$, while  $[X_j,X_k]=[P_j,P_k]=0$. Here we have introduced
$$
A_1=A+\frac{1}{2}\theta  B,\quad A_2=A-\frac{1}{2}\theta  B,\quad B_1=B-\frac{1}{2}\tilde\theta  A,\quad B_2=B+\frac{1}{2}\tilde\theta  A.
$$
The next step consists in introducing the following formally pseudo-bosonic operators: \be \left\{
    \begin{array}{ll} a_1=\frac{1}{2}\left(X_1+iP_1+iX_2-P_2\right),\quad
a_2=\frac{1}{2}\left(-iX_1+P_1-X_2-iP_2\right),    \\
b_1=\frac{1}{2}\left(X_1-iP_1-iX_2-P_2\right),\quad
b_2=\frac{1}{2}\left(iX_1+P_1-X_2+iP_2\right).
\end{array}
        \right.
        \label{65}\en
We see that $[a_j,b_k]=\delta_{j,k}\1$, while all the other commutators are zero, and that $b_j\neq a_j^\dagger$. We have used here the word
{\em formally} since we still have to check that Assumptions $\D$-PB 1, $\D$-PB 2 and $\D$-PB 3, are satisfied. In terms of these operators
$\hat H$ can be written as \be \hat H=(N_1+N_2+\1)+\frac{1}{2}(\theta+\tilde\theta)(N_1-N_2)+(A^2+B^2)\1.\label{66}\en Comparing this
hamiltonian with that is Example two, we see that the only difference is in the term $\frac{1}{2}(\theta+\tilde\theta)(N_1-N_2)$ which is
linear in the parameters $\theta$ and $\tilde\theta$.

As stated, so far ours are only formal computations. In order to make them rigorous, we have to check that the various assumptions of Section
II are satisfied. As usual, the first step consists in rewriting the operators $a_j$ and $b_j$ in terms of the variables $x_j$ and $p_j$. We
find:
$$\left\{
    \begin{array}{ll} a_1=\frac{1}{2}(x_1+ip_1+ix_2-p_2+k_1), \qquad a_2=\frac{1}{2}(-ix_1+p_1-x_2-ip_2+k_2),\\
b_1=\frac{1}{2}(x_1-ip_1-ix_2-p_2+\tilde k_1), \qquad b_2=\frac{1}{2}(ix_1+p_1-x_2+ip_2+\tilde k_2),\\
 \end{array}
        \right.$$
where
$$\left\{
    \begin{array}{ll} k_1=A\left(1-\frac{\tilde\theta}{2}\right)(i-1)+B\left(\frac{\theta}{2}-1\right)(i+1), \qquad
    k_2=A\left(1+\frac{\tilde\theta}{2}\right)(1-i)+B\left(\frac{\theta}{2}+1\right)(i+1),\\
\tilde k_1=A\left(1-\frac{\tilde\theta}{2}\right)(i+1)+B\left(\frac{\theta}{2}-1\right)(i-1), \qquad
\tilde k_2=-A\left(1+\frac{\tilde\theta}{2}\right)(i+1)+B\left(\frac{\theta}{2}+1\right)(i-1).
 \end{array}
        \right.$$
The function annihilated by $a_1$ and $a_2$ can be found easily, solving two coupled differential equations. The result is not very different
from what found in our previous examples:
$$
\varphi_{0,0}(x_1,x_2)=Ne^{-\frac{1}{2}(x_1^2+x_2^2)-\alpha_1 x_1-\alpha_2 x_2},\quad \Psi_{0,0}(x_1,x_2)=N'
e^{-\frac{1}{2}(x_1^2+x_2^2)+\alpha_1 x_1+\alpha_2 x_2},
$$
where $\alpha_1=\frac{k_1+ik_2}{2}$, $\alpha_2=\frac{k_1-ik_2}{2i}$, while $N$ and $N'$ are normalization constants, chosen in the usual way
(i.e. requiring that $\left<\varphi_{0,0},\Psi_{0,0}\right>=1$). Also in this model, it looks natural to take $\D\equiv\Sc({\Bbb R}^2)$. In
fact, with this choice, $\D$ is stable under the action of $a_j$, $b_j$, and their adjoints. Moreover, $\varphi_{0,0}(x_1,x_2)$ and
$\Psi_{0,0}(x_1,x_2)$ both belong to $\D$. The functions $\varphi_{n_1,n_2}(x_1,x_2)$ and $\Psi_{n_1,n_2}(x_1,x_2)$ are constructed in the
usual way, and they all belong to $\Sc({\Bbb R}^2)$. In conclusion,  Assumptions $\D$-pb 1 and $\D$-pb 2 are both satisfied.

\vspace{2mm}

For what concerns Assumption $\D$-pbw 3, the idea is again to look for an o.n. basis which is mapped into $\F_\varphi$ and $\F_\Psi$ by a
suitable operator. For that, it is convenient to introduce two bosonic operators, $c_j=\frac{1}{\sqrt{2}}(x_j+ip_j)$, $j=1,2$, and their
adjoints $c_j^\dagger$. Moreover, following \cite{messiah}, we introduce now two new bosonic lowering and raising operators
$c_g=\frac{1}{\sqrt{2}}(c_1+ic_2)$ and $c_d=\frac{-i}{\sqrt{2}}(c_1-ic_2)$. They satisfy the following:
$$
[c_g,c_g^\dagger]=[c_d,c_d^\dagger]=\1,
$$
all the other commutators being zero. The vacuum of $c_g$, $c_d$, $\chi_{0,0}$, coincides clearly with that of $c_1$, $c_2$, $\Phi_{0,0}$: in
other words, if  $c_1\Phi_{0,0}=c_2\Phi_{0,0}=0$, then, calling $\chi_{0,0}=\Phi_{0,0}$, we automatically have $c_d\chi_{0,0}=c_g\chi_{0,0}=0$.
Calling now $\chi_{n_d,n_g}=\frac{1}{\sqrt{n_d!\,n_g!}}(c_d^\dagger)^{n_d}(c_g^\dagger)^{n_g}\chi_{0,0}$, $n_d, n_g\geq0$, the set
$\F_\chi=\{\chi_{n_d,n_g}\}$ of all this vector is an o.n. basis for $\Hil$, \cite{messiah}. Introducing further the unbounded, self adjoint
and invertible operator $T$ as
$$
T=\exp\left\{-\frac{1}{2}\left(k_1c_g^\dagger+k_2c_d^\dagger+\overline{k_1}c_g+\overline{k_2}c_d\right)\right\},
$$
we find that
$$\left\{
    \begin{array}{ll} Tc_gT^{-1}=c_g+\frac{k_1}{2}=a_1, \qquad
    Tc_dT^{-1}=c_d+\frac{k_2}{2}=a_2,\\
Tc_g^\dagger T^{-1}=c_g^\dagger-\frac{\overline{k_1}}{2}=b_1, \qquad
    Tc_d^\dagger T^{-1}=c_d^\dagger -\frac{\overline{k_2}}{2}=b_2.
 \end{array}
        \right.$$
Now, except at most for an unessential normalization, we can check that $\varphi_{0,0}=T\chi_{0,0}$ and that $\Psi_{0,0}=T^{-1}\chi_{0,0}$.
This follows, for instance, from the fact that $c_g(T\chi_{0,0})=-\frac{k_1}{2}(T\chi_{0,0})$ and
$c_d(T\chi_{0,0})=-\frac{k_2}{2}(T\chi_{0,0})$. These equalities can be now easily extended to all the vectors:
$\varphi_{n_1,n_2}=T\chi_{n_1,n_2}$ and $\Psi_{n_1,n_2}=T^{-1}\chi_{n_1,n_2}$, for all $n_j\geq0$. This allow us to use Proposition
\ref{propa1}, so that we can conclude that $\F_\varphi$ and $\F_\Psi$ are $D(T)\cap D(T^{-1})$-quasi bases for $\Hil$.
Needless to say, the same final construction can again be repeated. In particular, the operator $\Theta$ can be introduced, mapping $\F_\Psi$
into $\F_\varphi$ and viceversa.

\section{Conclusions}

We have shown how several quantum mechanical systems, recently proposed in the context of PT-quantum mechanics, fits in our general
pseudo-bosonic settings. In this way some of those aspect which were not considered in \cite{ben1,miao1,miao2}, for instance the construction
of the eigenstates of $H^\dagger$, can be naturally discussed.

Our feeling is that, whenever we have to do with a non self-adjoint hamiltonian whose eigenvalues are linear in the quantum numbers needed to
describe the system, $\D$-PB may be the right objects to introduce in the model. We hope to be able to transform this feeling in a formal
theorem soon. This might have interesting applications in other concrete physical systems, mainly those recently introduced in quantum optics
and, more in general, in gain-loss structures.

\section*{Acknowledgements}

This work was partially supported by the University of Palermo.

\end{document}